\documentclass[12pt,preprint]{aastex}
\usepackage{graphicx}

\begin{document}

\title{The Chandra HRC View of the Sub-arcsecond Structures in the Nuclear Region of NGC~1068}

\author{Junfeng Wang\altaffilmark{1}, Giuseppina Fabbiano\altaffilmark{1}, Margarita Karovska\altaffilmark{1}, Martin  Elvis\altaffilmark{1}, Guido Risaliti\altaffilmark{1,2}}

 \altaffiltext{1}{Harvard-Smithsonian Center for Astrophysics, 60 Garden St, Cambridge, MA 02138}
 \altaffiltext{2}{INAF-Arcetri Observatory, Largo E, Fermi 5, I-50125 Firenze, Italy}

\email{juwang@cfa.harvard.edu}

\begin{abstract}

We have obtained a high spatial resolution X-ray image of the nucleus
of NGC 1068 using the High Resolution Camera (HRC-I) on board the {\em
  Chandra X-ray Observatory}, which provides an unprecedented view of
the innermost 1 arcsec radius region of this galaxy.  The HRC image
resolves the narrow line region into X-ray emission clumps matching
bright emission-line clouds in the HST [OIII] $\lambda$5007 images and
allows comparison with sub-arcsec scale radio jet for the first time.
Two distinct X-ray knots are revealed at 1.3-1.4 arcsec northeast and
southwest of the nucleus.  Based on the combined X-ray, [OIII], and
radio continuum morphology, we identify the locations of intense radio
jet--cloud interaction.  The [OIII] to soft X-ray ratios show that
some of these clouds are strongly affected by shock heating, whereas
in other locations the jet simply thrusts through with no signs of
strong interaction. This is further strengthened by the presence of a
$kT\sim 1$ keV collisionally ionized component in the ACIS spectrum of
a shock heated cloud HST-G.  We estimate that the kinematic luminosity
of the jet-driven shocks is $6\times 10^{38}$ erg s$^{-1}$, a
negligible fraction ($10^{-4}$) of the estimated total jet power.

\end{abstract}


\section{Introduction}

NGC 1068 is the brightest prototype Seyfert 2 active galactic nucleus
(AGN), and also one of the closest, with a distance of only 14.4 Mpc
\citep[$1\arcsec=72$~pc;][]{Bland-Hawthorn97}. This makes NGC 1068 an
optimal target to investigate the morphology and kinematics of the
different components in the AGN central regions: the infrared-emitting
molecular gas, the X-ray obscuring torus, and the ionized gas
\citep[e.g.,][]{Exposito11,Krips11,SB12}.

Several detailed studies have been performed at multi-wavelength with
the highest spatial resolution instruments available, probing the
nuclear region down to scales of the order of, or smaller than
$\sim$10 pc \citep[e.g.,][]{Evans91,Capetti97,Gallimore04,Groves04}.
The picture emerging from these studies is that of a complex
circum-nuclear environment \citep{Raban09}.  Previous X-ray studies
identified a compact photoionized component, unresolved on scales of
$\sim$100 pc, emitting the soft X-ray spectrum studied through
dispersive grating spectroscopy \citep{Kinkhabwala02,Ogle03,Evans10},
and an ionization cone on $\sim$kpc scales, observed in HST [OIII]
images, broadly coincident with an extended X-ray emission region
imaged with {\em Chandra} ACIS \citep{Young01,Ogle03}.

Besides reporting a general good correspondence between the [OIII]
distribution and the X-ray morphology, \citet{Young01} found that the
ACIS-S spectra of several diffuse emission regions could not be fitted
with thermal plasma models. \citet{Brinkman02} and \citet{Ogle03}
demonstrated that the X-ray emission of the centermost region
(unresolved with ACIS) and that of a cloud NE of the nucleus is
dominated by photoionization. This would also explain the general
agreement between X-ray and [OIII] morphology.

While these conclusions are strong, we may be still missing a part of
the story. In particular, the radial velocities of the emission lines
suggest forces at work within 6\arcsec\/ of the nucleus, accelerating
outflows in the central ~2\arcsec\/ and then gradual deceleration
\citep{Crenshaw00}. These motions suggest clouds accelerated by
nuclear winds, then decelerating because of interaction with the
ambient ISM; these velocities may also reflect the interaction of the
radio jets with the ISM \citep{Axon98,Ogle03}.  The inner few hundred
parsec region is where jet-ISM interactions are expected to be most
intensive, but this region has not been investigated in great detail
in X-rays.  Previous deep {\em Chandra} ACIS observations, while providing
illuminating X-ray findings, suffered significant photon pileup in the
$r\leq 2\arcsec$ ($\leq 144$~pc) nuclear region. Even the extended
emission is affected by pileup in the ACIS exposure with a 3.2 s frame
time \citep{Young01}.

In this paper, we report on the X-ray morphology mapped with a
resolution of $0.26\arcsec$, $\sim 22$ pc, in the nuclear region
obtained from the deconvolved HRC imaging of NGC 1068, which is free
from pileup effect.  Taking advantage of archival ACIS image with
subpixel resolution, we are able to identify X-ray structures with
features seen in other wavebands and locations of intense jet-cloud
interaction.  Detailed imaging and spectral analysis of X-ray emission
from the nucleus to galaxy scale will be presented in a forthcoming
paper, utilizing all available {\em Chandra} ACIS and HRC data.

\section{Observations and Data Reduction}

NGC 1068 was observed on October 2, 2010, with the {\em Chandra} HRC-I
\citep{Murray00} for 40.18 ks (ObsID 12705, PI: Fabbiano).  The
nominal pointing was ($\alpha=02^{\rm h}42^{\rm m}41.^{\rm s}28$,
$\delta= -00^{\circ}00^{\prime}34.^{\prime \prime}9$) with 82.9 degree
roll angle.  The HRC data were
reprocessed\footnote{\url{http://cxc.harvard.edu/ciao/threads/createL2/}}
with the {\em Chandra} Interactive Analysis of Observations software
package (CIAO; version 4.2) tool {\tt
hrc\_process\_events}\footnote{\url{http://cxc.harvard.edu/ciao4.2/ahelp/hrc\_process\_events.html}}
using {\em Chandra} Calibration Database (CALDB) version 4.3.1.  The
net exposure time was 39.7 ks after screening for brief periods of
elevated background.  Since there is a known
artifact\footnote{\url{http://cxc.harvard.edu/ciao4.4/caveats/psf\_artifact.html}}
in the {\em Chandra}/HRC point spread function (PSF), we have used the CIAO
tool {\tt make\_psf\_asymmetry\_region} to identify its location as
precaution.  The artifact is in the SE quadrant (position angle
[P.A.]$\sim$87$^{\rm o}$ to $\sim$137$^{\rm o}$), not related to any
of the morphological features in NGC 1068 that we resolve and discuss
in later sections.  The archival ACIS S3 observation \citep[ObsID 370,
previously presented in][]{Young01} shown in this work was taken on
February 22, 2002 with 0.1 s frame time and a total exposure of 1.43
ks.  We reprocessed the ACIS data to generate new level 2 file that
has the latest calibration and the subpixel resolution algorithm
\citep[``EDSER'',][]{Li03} applied.  To evaluate the accuracy of
astrometry, we compared the positions of X-ray point sources to the
coordinates from the the Naval Observatory Merged Astrometric Dataset
\citep[NOMAD;][]{Zacharias04} and derived absolute astrometric
accuracy of $0.3\arcsec$ (1$\sigma$).

\section{Imaging and Spectral Analysis}

Following our {\em Chandra} HRC study of the Seyfert 1 galaxy NGC 4151
\citep{Wang09a}, we performed image restoration using the expectation
through Markov Chain Monte Carlo (EMC2) algorithm
\citep{Esch04,Karovska05}.  The effectiveness of this method was
demonstrated with a number of astronomical imaging studies
\citep{Karovska05, Karovska07, Wang09a}. The {\em Chandra} PSF was
simulated with the {\em Chandra} Ray Tracer
(ChaRT\footnote{\url{http://cxc.harvard.edu/chart/}}) using a
monochromatic energy at 1 keV (recommended for HRC data; see ChaRT
thread noted above).  The rays were then projected onto the HRC
detector with CIAO tool {\tt psf\_project\_ray} adopting a
$0.2\arcsec$ Gaussian blurring to generate the PSF image.

The EMC2 restored (500 iterations) HRC image of the NGC 1068 nuclear
region (the central $8^{\prime\prime}\times 8^{\prime\prime}$) is
presented in Figure~\ref{fig1}, which also shows the raw HRC-I image,
and the ACIS-S images at native pixel resolution and at subpixel
resolution.

\subsection{X-ray Morphology of the Narrow-Line Region}

Although all four {\em Chandra} images show similar northeast--southwest
elongation, finer features are seen in the subpixel rebinned ACIS
image, which were not discernible in the raw ACIS image due to
undersampling with the native pixels \citep[c.f. Figure 2
in][]{Young01}. These higher resolution features are also evident in
the raw HRC image and more strikingly in the PSF-deconvolved HRC
image.

The most probable location of the nucleus is generally accepted to be
the VLBI radio source ``S1'' \citep{Gallimore96a, Kishimoto99}.  We
registered our X-ray image so that the X-ray peak emission is
coincident with UV peak in \citep{Capetti97}, and 0.1\arcsec\/ north
of S1 (the active nucleus), consistent with an indirect view of the
nucleus in the X-rays \citep[obscured by a column density exceeding
$10^{25}$ cm$^{-2}$,][]{Matt00}.

\begin{figure}
%\plottwo{HRC_7arcsec.ps}{HRC_decon_7arcsec.ps}
\centerline{ \includegraphics[scale=0.45]{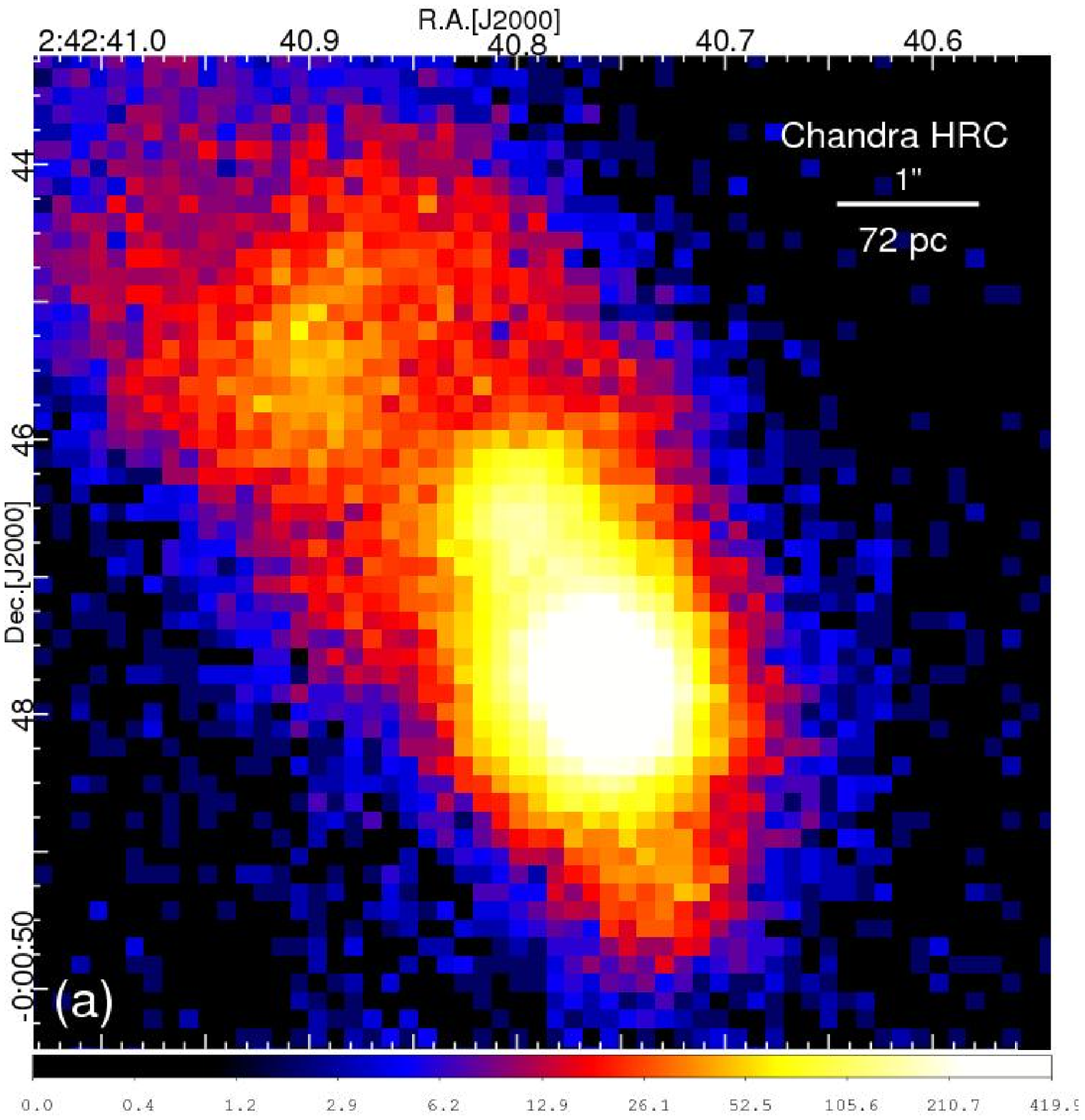} \includegraphics[scale=0.45]{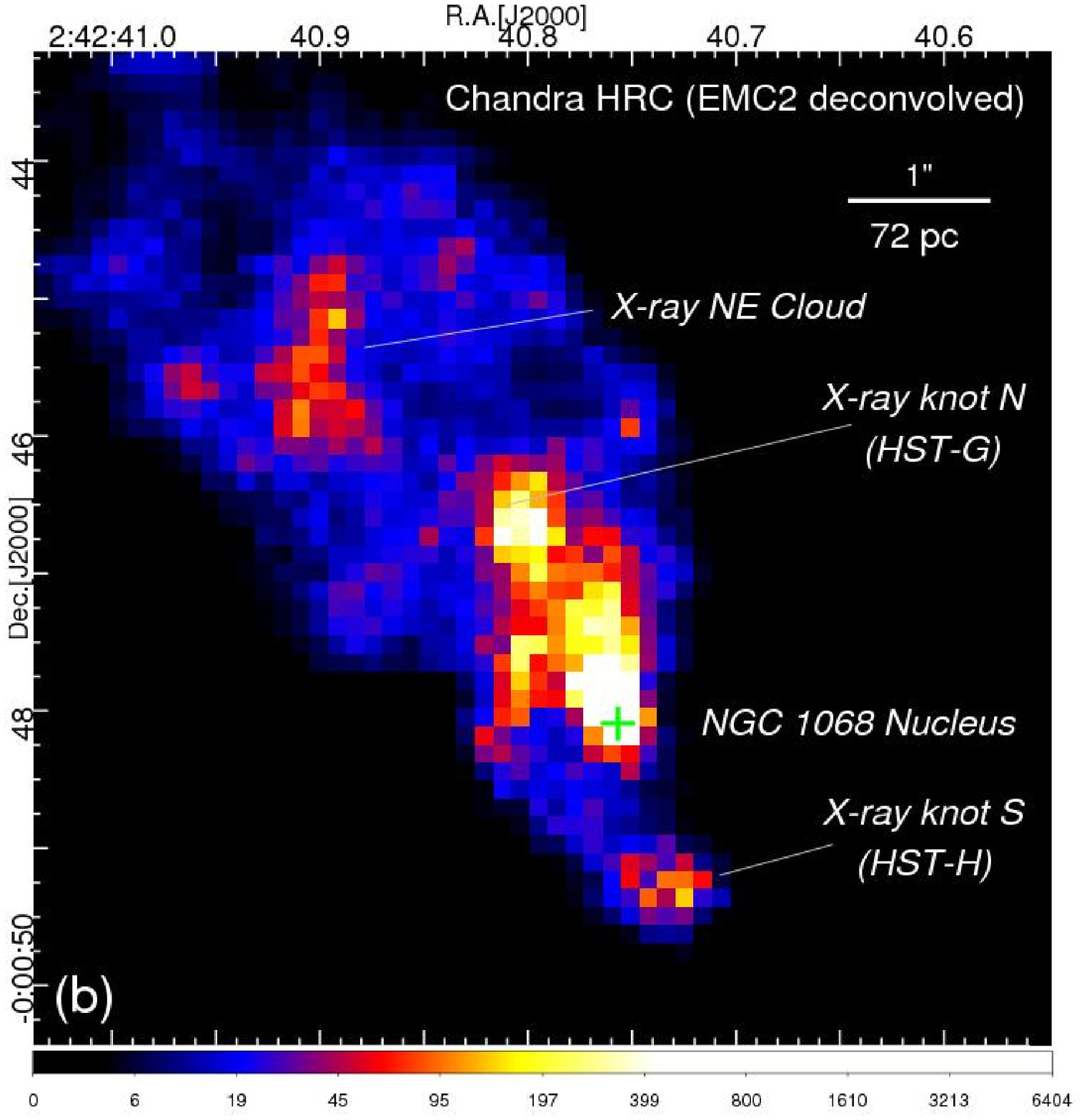}}
\centerline{ \includegraphics[scale=0.45]{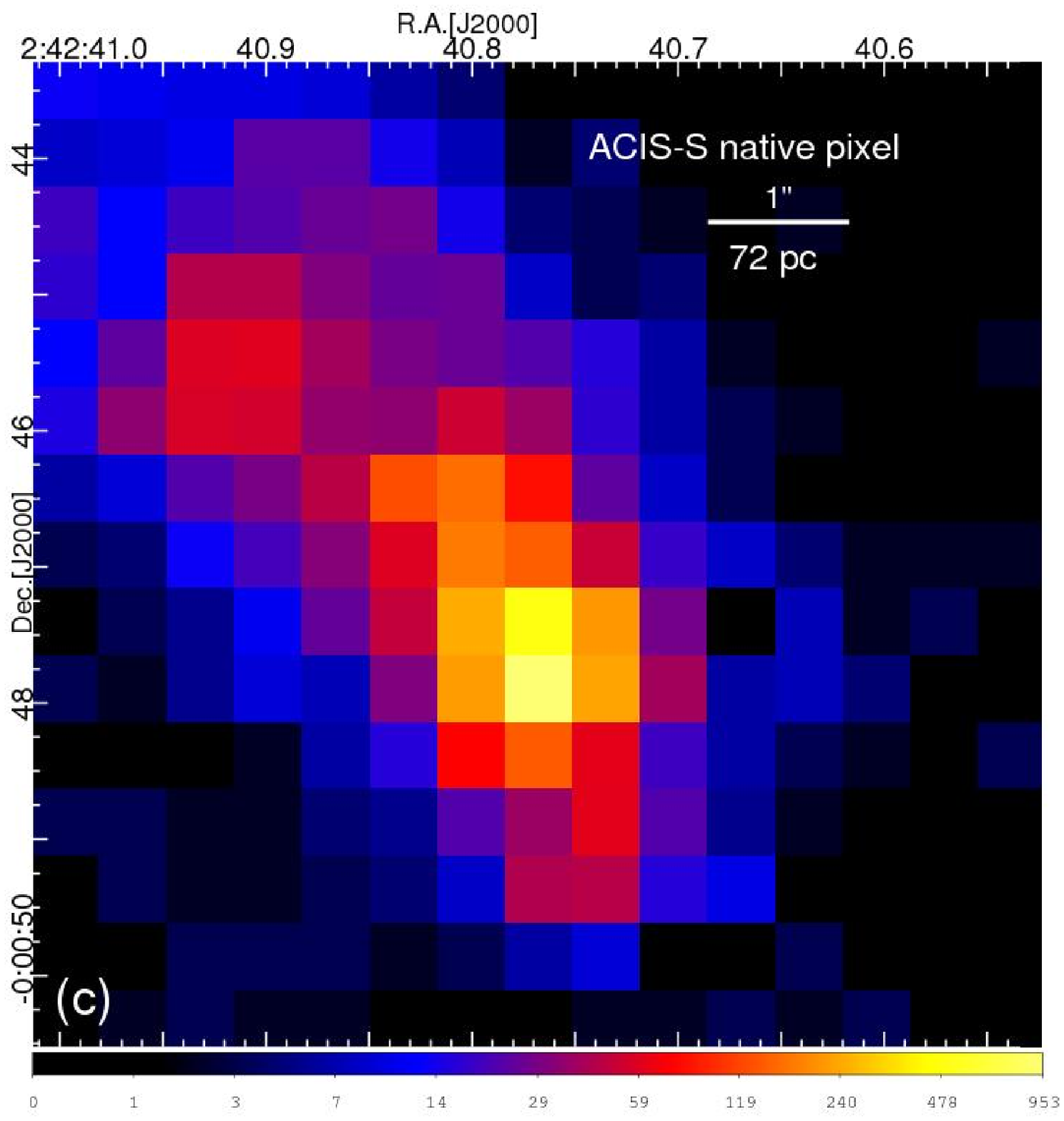} \includegraphics[scale=0.45]{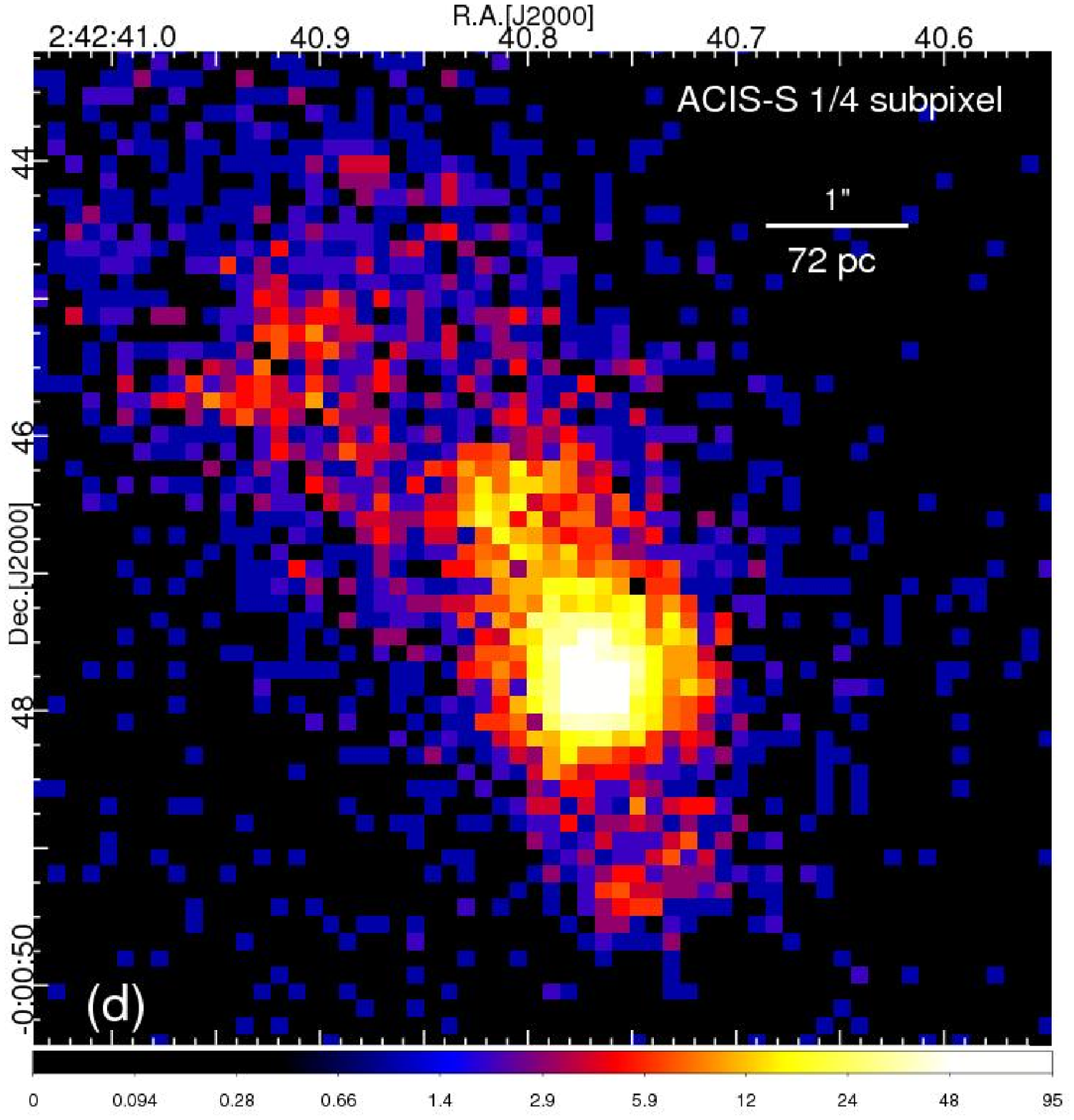}}
\caption{{\footnotesize (a) The HRC image of the 7 arcsec across (500 pc) nuclear
  region of NGC 1068; (b) EMC2 deconvolved HRC image; (c) The ACIS 0.3--2 keV image (0.1 s frame time) of the same region, displayed at the native pixel (0.492\arcsec\/); (d) The subpixel repositioned, rebinned ACIS image of the same region displayed at 1/4 of the native pixel (0.123\arcsec\/). A 1\arcsec\/ scale bar is shown in all panels, corresponding to 72 pc at the distance of NGC 1068.}
\label{fig1}}
\end{figure}

The X-ray emission in the northeast (NE) direction, where the optical
extended narrow line region is, shows a cylindrical morphology with
regions of apparent X-ray enhancement (``clumps'') and deficit
(``bubbles'').  The previously identified X-ray emitting ``NE cloud''
\citep[$\sim$3\arcsec\/ NE of the nucleus;][]{Young01, Ogle03} is
clearly resolved.  Here we focus on the inner $\sim$200 pc
(3\arcsec\/) scale region that has been poorly studied in the X-rays
before and now becomes accessible with the HRC deconvolved image
(Figure~\ref{fig1}b).  The subpixel resolution ACIS image of 0.1 s
frame time was not used for deconvolution, because the appropriate
model for the energy dependent, subpixel resolution PSF is not yet
available and there is still low level ($\sim$5\%) pileup in the
nucleus ($3.32\pm0.048$ counts s$^{-1}$).

The X-ray emission in the nuclear region is resolved into distinct
components in the HRC image, namely a bright nucleus with elongation
to the north, and two blobs $\sim$0.\farcs\/7 from the nucleus at
P.A.$\sim$6$^{\rm o}$ and P.A.$\sim$54$^{\rm o}$, respectively.  A
bright knot of X-ray emission is uncovered at 1.4\arcsec\/ to the NE
of the nucleus at P.A.$\sim$25$^{\rm o}$ (labelled as ``X-ray knot N''
in Figure~\ref{fig1}b, and throughout the paper referred as ``HST-G''
following HST features in the optical and avoid inventing new
nomenclature, see next section).  A similar but fainter feature is
located at 1.3\arcsec\/ SW of the nucleus (X-ray knot S, or
``HST-H'').  These two knots are also clearly present in the ACIS
subpixel image (Figure~\ref{fig1}d), which shares a great similarity
of morphology with the HRC images.

\subsection{Comparison with High Resolution [OIII] Image and Radio Maps}

\begin{figure}
\centerline{ \includegraphics[scale=0.43]{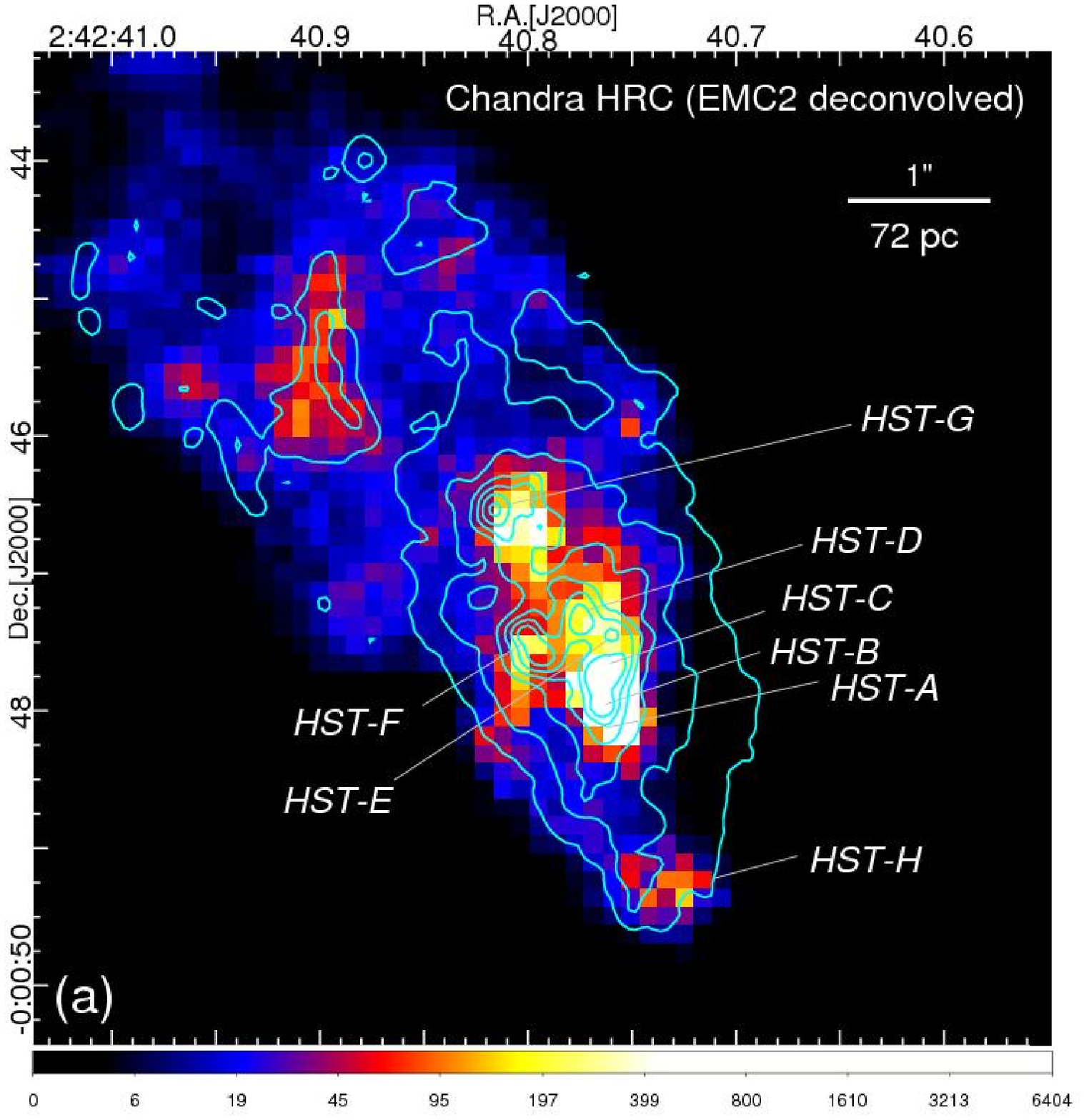} \includegraphics[scale=0.4]{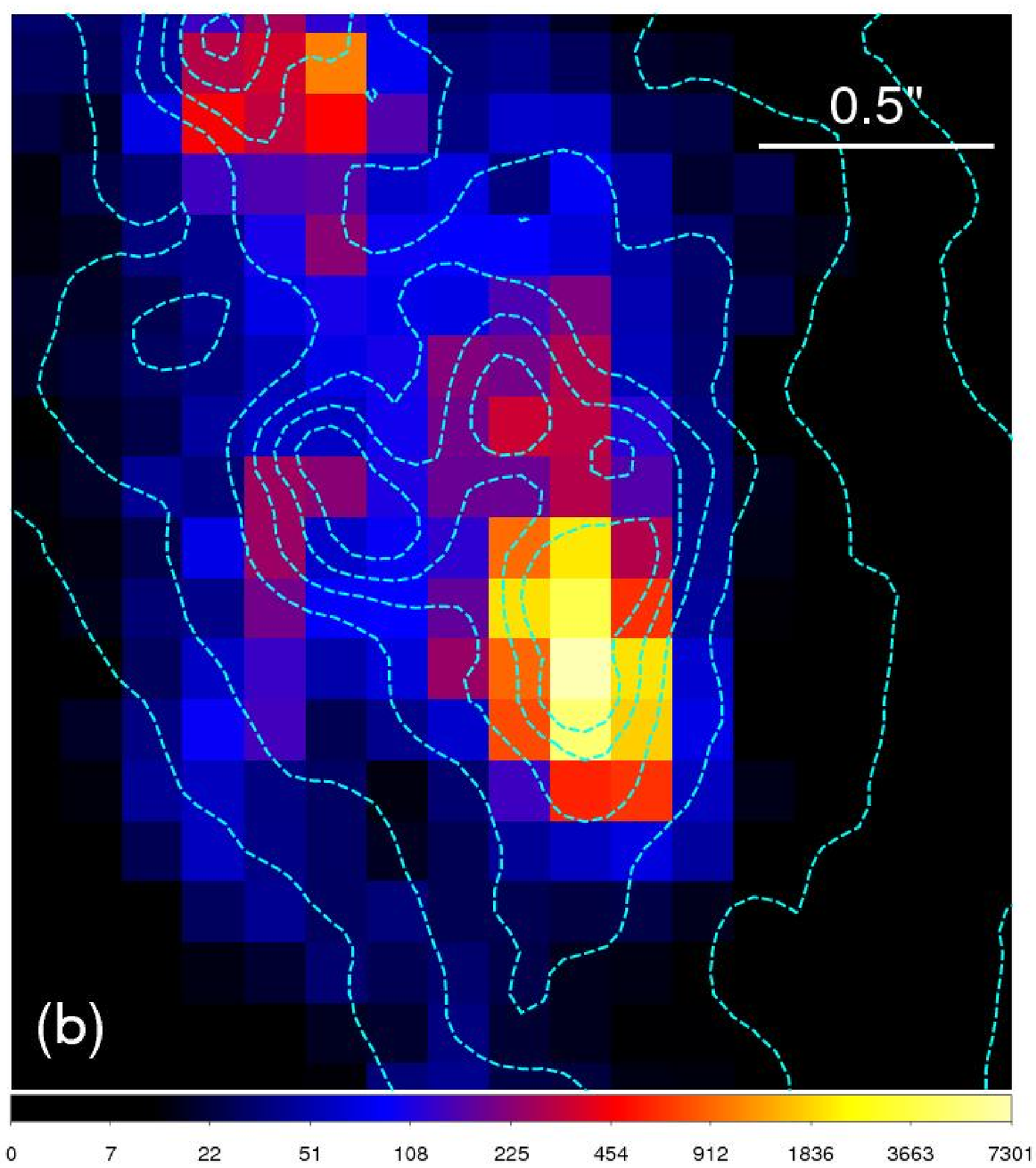}}
\centerline{ \includegraphics[scale=0.43]{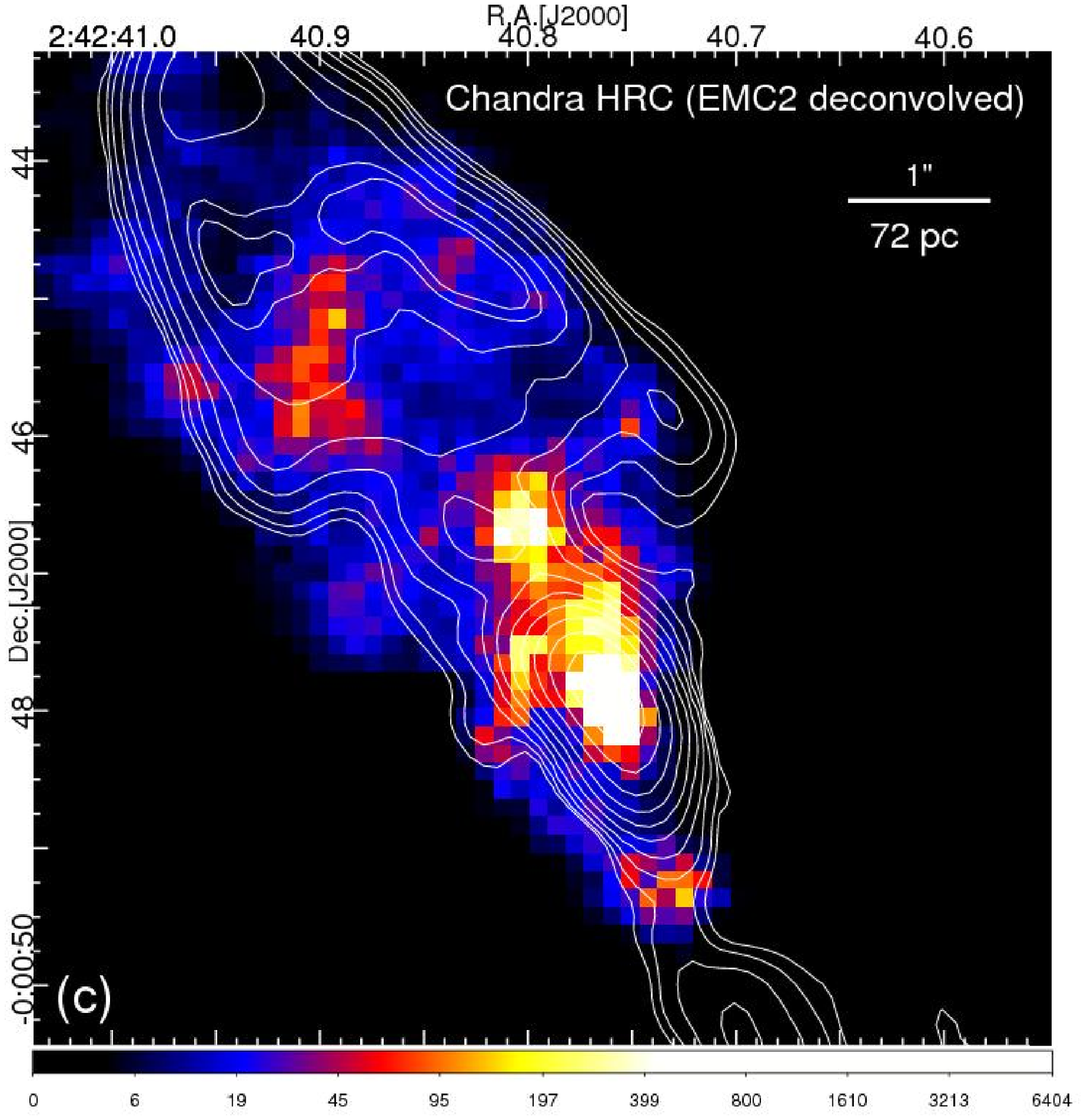} \includegraphics[scale=0.4]{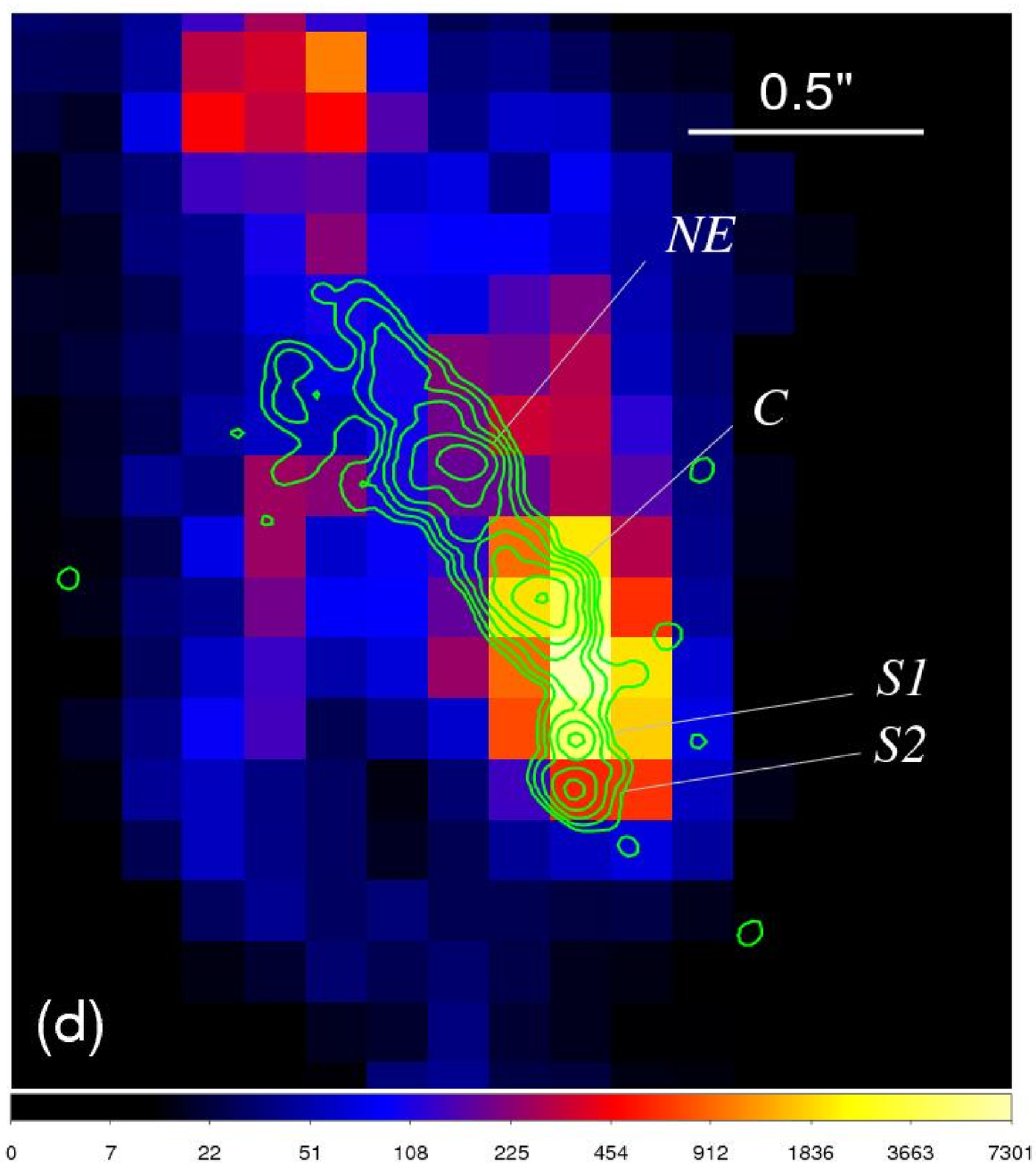}}
\epsscale{1.0}
\caption{{\footnotesize (a) The deconvolved HRC image of the X-ray
    emission in NGC 1068 overlaid with the contours of HST/WFPC2
    [OIII] $\lambda$5007 line emission (Capetti et al.\ 1997). Contour
    levels are from $3\times 10^{-16}$ to $3\times 10^{-14}$ ers
    cm$^{-2}$ s$^{-1}$ (PC pixel)$^{-1}$. (b) Zoom-in image showing
    the detailed correspondance of [OIII] and X-ray features in the
    innermost region; (c) The deconvolved HRC image overlaid with the
    contours of VLA 5 GHz emission (Wilson \& Ulvestad 1983). The beam
    size is 0\farcs\/49$\times$0\farcs\/38.  Contour levels are
    0.3, 0.7, 1.4, 2.8, 5.6, 16, 27.9, 59.5, 120, and 224 mJy
    beam$^{-1}$. (d) The innermost region overlaid with the MERLIN 6
    cm contours (Gallimore et al.\ 1996). The beam size is
    0\farcs\/065. Contour levels are 0.862, 1.73, 3.50, 7.06, 14.23,
    28.7, and 57.8 mJy beam$^{-1}$. Note the 0\farcs\/5 scale bar
    in the zoom-in panels (b) and (d).}
\label{fig2}}
\end{figure}

We further compare the rich structures in [OIII]$\lambda$5007 emission
line and radio images to the X-ray emission.  Figures~\ref{fig2}a
and~\ref{fig2}b show the HRC image overlaid with contours of the
[OIII] emission from the {\em HST} WFPC2/F502N image presented in
\citet{Capetti97} and the innermost zoom covers 150 pc region (2
arcsec across). The labelled features following the HST clouds naming
convention in \citet{Evans91}. Previous ACIS study already showed a
general spatial correlation between the two wavebands \citep{Young01},
the HRC image firmly demonstrates that there is remarkable
morphological agreement between the X-ray clumps and the
[OIII]$\lambda$5007 emission, even at the smallest spatial scales.

A similar comparison with the VLA A-array 6 cm map \citep{Wilson83,
Gallimore96a} is shown in Figure~\ref{fig2}c.  We find that the X-ray
``NE cloud'' \citep{Ogle03} does not correspond to any enhancement in
the radio lobe, although it is slightly offset (1\arcsec\/ to the SW)
from a bright radio filament. The newly identified X-ray knot (HST-G)
is 0.2\arcsec\/ from the peak of a radio knot, where the collimated
radio outflow becomes more diffuse and lobe-like.

The zoom (Figure~\ref{fig2}d) shows fine details of the spatial
relations with the MERLIN 6 cm features \citep[components NE, C, S1,
and S2 in][]{Gallimore96a}.  The X-ray elongation northward of the
nucleus follows the linear radio emission between S1 and C, then
become faint and spatially anti-correlated with the radio emission
after the jet shows significant ``bending'' at component C
\citep{Gallimore96a}.

\subsection{X-ray Spectral Modeling of HST Cloud G}

\begin{figure}
\includegraphics[scale=0.6,angle=-90]{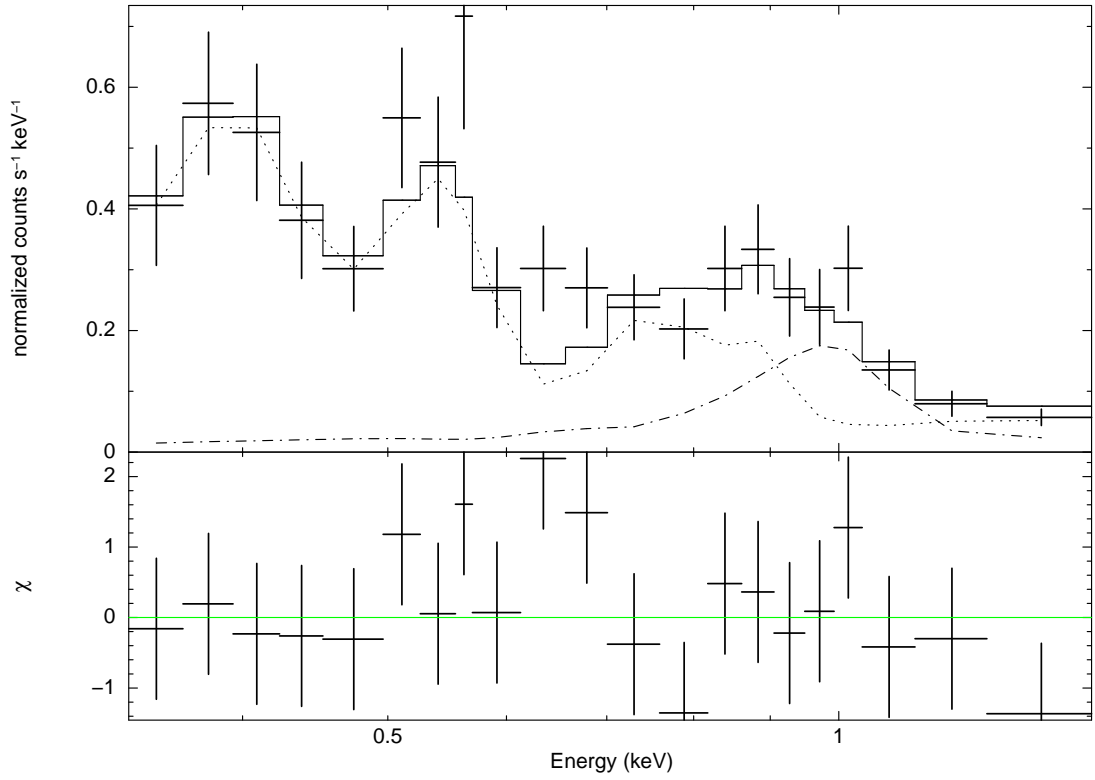}
\caption{The X-ray spectrum of the cloud HST-G, and the best fit model
  consisting of a $kT=1.08$ keV thermal component (the dash-dotted line) and a low ionization
  photoionized emission component (the dotted line).
\label{fig3}}
\end{figure}

The 0.1 s frame time ACIS image with subpixel resolution clearly
resolves HST-G and HST-H (Figure~\ref{fig1}d), which are free of
pileup.  Because of the short exposure (1.4 ks), only the brighter
HST-G have enough counts ($386\pm 19$) to attempt spectral fit using
photoionization models.  The X-ray spectrum was extracted using CIAO
tool {\tt specextract} from a $0.5\arcsec$-radius circle centered on
the X-ray knot, which was grouped to have a minimum of 15 counts per
bin to allow usage of $\chi^2$ statistics.  Spectral modeling was
performed following \citet{Wang11b} with the XSPEC package (version
12.7.1) and photoionization model grids generated by {\tt Cloudy}
\citep[][version 08.01]{Ferland04}.  A single ionization
($logU=-1.7$,$logN_H=21.7$) component is not adequate to obtain a
reasonable fit to the spectrum (reduced $\chi^2=2.2$).  The addition
of an optically thin thermal plasma component (APEC) with the best fit
temperature of $kT=1.08^{+0.21}_{-0.12}$ keV significantly improves
the fit ($\chi_{\nu}^2=1.1$, $\Delta \chi^2=22$; $F$-statistic=9.94 or
probability of improvement 99.8\%).  There is a residual line feature
at $\sim 0.65$ keV (see Figure~\ref{fig3}) that is probably due to
OVIII Ly$\alpha$, and can be fit with an additional high ionization
component ($logU=0.9$, $\chi_{\nu}^2=0.8$).  The probability of
improvement over current model is 90\%.  However, this has little
effect on the thermal component. The flux contributed by the thermal
component is $F_{0.5-2 {\rm keV}}=1.8\pm 0.6\times 10^{-13}$ erg
s$^{-1}$ cm$^{-2}$ ($L_{0.5-2{\rm keV}}=4.5\times 10^{39}$ erg
s$^{-1}$), which is approximately 30\% of the total X-ray emission
from cloud G.

\section{Discussion}

The unprecedented high spatial resolution HRC image of NGC 1068 has
enabled us to compare the sub-arcsecond structures of the X-ray
emission and those of the HST NLR clouds in the nuclear region, where
the ACIS data is affected by pileup.  Overall, there is a good
correlation between enhancements in [OIII] and X-ray emission
(Figure~\ref{fig2}a,b) in the nuclear region of NGC 1068, where the
grating spectroscopic observations resolve the soft X-ray emission
into many lines characterized by photoionization and photoexcitation
(e.g., NVII, OVII, OVIII, and NeIX), consistent with previous
suggestions that both originate from gas photoionized by the AGN
\citep[e.g.,][]{Young01, Brinkman02, Ogle03}.  This general
morphological agreement appears to be common for kpc-scale soft X-ray
and [OIII] emission in nearby Seyfert galaxies
\citep[e.g.,][]{Bianchi06, Bianchi10, Wang11a, Paggi12}.  The ratios
between [OIII] and soft X-rays were found to effectively trace the
ionization states \citep[e.g.,][]{Bianchi06, Wang09a}.  Following
these studies, we use [OIII]/soft X-ray ratios of different clouds to
locate sites of shocks produced in strong jet--cloud interaction, as
indicated by low [OIII]/soft X-ray ratios due to enhanced thermal
X-ray emission from shock heating \citep{Bicknell98, Wang09a,
Wang11b}.

\begin{figure}
\epsscale{1.1}
\plotone{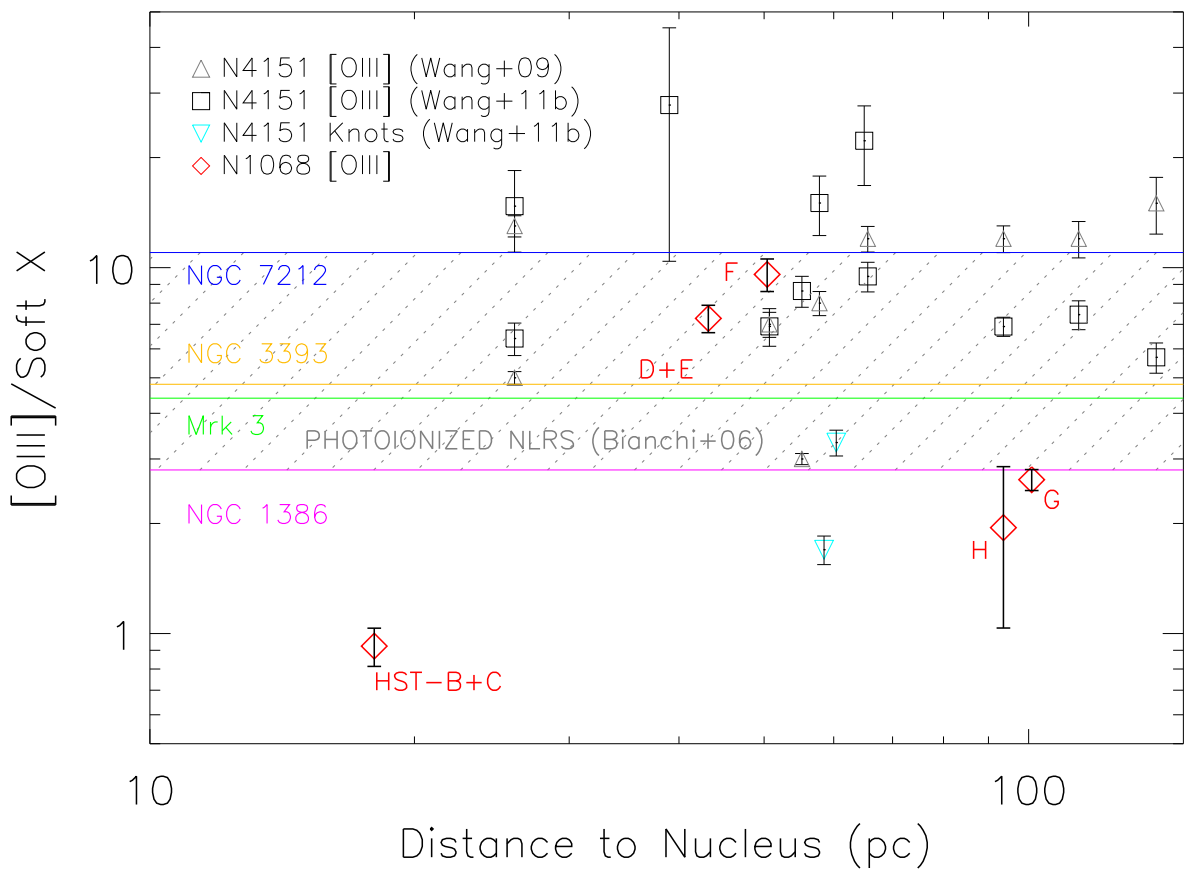}
\caption{The [OIII] to soft X-ray ratio as a function of the cloud's
  distance to the nucleus.  The (red) diamonds are the clouds as
  labeled in Figure~\ref{fig2} and Table~\ref{flux}, and previous
  measurements for NGC 4151 clouds (Wang et al. 2009a; Wang et
  al. 2011b) are also shown for reference.  The shaded area is the
  range of ratios in Bianchi et al. (2006) that can be explained with
  photoionization modeling. The blue, gold, green, and magenta lines
  indicate the [OIII]/X-ray ratios for the NLRs of NGC 7212, NGC 3393,
  Mrk 3, and NGC 1386 (Bianchi et al. 2006), respectively.
\label{fig4}}
\end{figure}

Using the calibrated WFPC2 F502N image \citep{Capetti97}, we measured
the [OIII] fluxes for the clouds (Evans et al. 1991) following WFPC2
narrow-band photometry in WFPC2 Data
Handbook\footnote{\url{http://documents.stsci.edu/hst/wfpc2/documents/handbooks/dhb/wfpc2\_cover.html}}. The
0.5--2 keV X-ray fluxes were measured from counts extracted from the
same regions using the HRC image.  The central positions (right
ascension and declination in J2000) of these clouds and the
rectangular regions used to estimate the [OIII] and X-ray fluxes were
listed in Table~\ref{flux}.  It should be noted that there are finer
[OIII] features resolved in the HST image that are not resolved in the
X-rays, so these fluxes only reflect the average values for the
ionized gas.  We assumed the phenomenological model (two-component
bremsstrahlung plus power law) for the X-ray emission ($kT=0.45$ keV
and $\Gamma=1.0$, $N_H=2.99\times 10^{20}$ cm$^{-2}$; Young et
al. 2001) and obtained $F_{0.5-2keV}=5.2\times 10^{-12}$ erg s$^{-1}$
cm$^{-2}$ per HRC count s$^{-1}$ for the flux conversion.  In
Figure~\ref{fig4} we show the [OIII] to soft X-ray ratio for five
distinct cloud features along the jet (Table~\ref{flux}) at various
radii to the nucleus ($\sim$20 pc--100 pc).

We note that three regions (HST-B+C, HST-G, and the previously
unlabeled HST-H) show significantly lower [OIII]/X-ray ratio ($\sim$1)
than the typical range spanned by photoionized clouds
\citep[$\sim$3--11 in the Seyfert galaxies observed by][]{Bianchi06},
implying higher collisional ionization at these locations.  We suggest
that these clouds are strongly interacting with the jet, and that
shocks driven into the obstructing clouds produce thermal X-ray
emission.  Indeed, our spectral fitting of cloud G revealed presence
of $kT=1$ keV hot plasma.  The remaining two regions (HST-D+E and
HST-F), which are clouds that bracket the MERLIN knot ``NE''
\citep[][Figure~\ref{fig2}d]{Gallimore96b}, show no evidence of
jet-cloud collision.  This scenario is consistent with the
multi-wavelength morphology of the circum-nuclear region.  At the
radio knot C, the P.A. of the jet changes from $\sim 11^{\rm o}$ to
$\sim 33^{\rm o}$ \citep{Gallimore96a}; \citet{Gallimore96b} suggested
that it likely results from shock interaction between the radio jet
and a dense molecular cloud, which was indeed identified in 2.12$\mu$m
H$_2$(1-0) S1 emission \citep{Muller-Sanchez09}.  Similarly, at the
clouds HST-G and HST-H, the jet changes into a lobe after driving
shocks into these clouds, as shown by our observations.  There are
other regions also show low [OIII]/X-ray ratios, such as the region
between HST-D and HST-G, and a narrow ``bar'' feature (NW-SE) at the
base of the radio lobe.  These regions are likely associated with
collisional gas as well, although much fainter and diffuse to measure
the [OIII] emission.

The kinematics of the ionized gas also suggest jet-cloud impact.  The
long slit spectroscopy of NGC 1068 with HST FOC \citep{Axon98} found
that the NLR gas at HST-G is strongly kinematically perturbed and
shows velocity systems separated by 1500 km
s$^{-1}$. \citet{Crenshaw00} show evidence for acceleration of the
radio outflow from the nucleus and subsequently deceleration to
systemic velocity.  The same behavior was seen in the high ionizing
potential coronal line region (CLR) gas \citep{Muller-Sanchez11},
where velocity tomography of coronal lines like [Si VI], suggested a
turnover point for the radial acceleration at 80 pc NE from the
nucleus, at the position of cloud HST-G, where the jet changes into a
lobe.  Moreover, the detection of extended highly ionized gas (e.g.,
[Fe VII]) in NGC 1068 led \citet{Rodriguez06} to the suggestion that
shocks related to the outflow could be required to power the line
formation.

Last but not least, the spectral modeling of cloud HST-G revealed the
presence of $kT\sim 1$ keV hot gas.  Assuming a spherical symmetry for
the gas ($r=0.5\arcsec$ or 36 pc), we can derive the emission measure
and electron density for the hot gas \citep[e.g.,][]{Wang09b}. Of most
interest, the estimated thermal energy content deposited by the
jet-ISM interaction is $E_{thermal}=1.9\times 10^{53}$ ergs.  Taking
the velocity of the fast shocks driven by the jet to be $V\simeq 700$
km s$^{-1}$ \citep{Axon98}, the relevant timescale is the crossing
time for the shocks to move through the 1\arcsec\/ (72 pc) extent of
the cloud, $t_{cross}=d/V=10^5$ yr.  If the jet-cloud interaction
converts kinematic energy into heating of the hot gas, this implies a
kinematic luminosity (or energy injection rate) of $L_{K.E.}=6\times
10^{38}$ erg s$^{-1}$, which is lower but comparable to the kinematic
luminosity measured from the jet-cloud interaction in NGC 4151
($L_{K.E.}=1\times 10^{39}$ erg s$^{-1}$).  To put this in the context
of the available jet power, we derive the total jet power of NGC 1068,
$P_{jet}\sim 7\times 10^{42}$ erg s$^{-1}$, adopting the flux density
of $4.8\pm 0.2$ Jy at 1.465 GHz \citep{Wilson83} and the
$P_{jet}$--$P_{radio}$ relation from a sample of radio loud galaxies
with X-ray cavities \citep{Cavagnolo10}.  A negligible amount of the
jet power ($L_{K.E.}/P_{jet}=10^{-4}$) is lost to the ISM during the
encounter with the cloud G.
 
Our {\em Chandra} HRC and ACIS results show three locations of intense
interaction between the radio jet in NGC 1068 and the NLR clouds in
the inner 200 pc: the jet first encounters a dense clump at knot C
(HST-B+C), produces a shock front \citep{Gallimore96a,Gallimore96b}
and changes direction; then it sweeps through between clouds D+E and F
with no evidence of interaction; it interacts again with clouds HST-G
in the NE and HST-H in the SW, changing morphology and ionization
structure \citep{Capetti97}.

\section{Conclusions}

We have obtained high spatial resolution X-ray image of the nucleus of
NGC 1068 using the HRC, which allows, for the first time, a direct
view of the innermost 1 arcsec (72 pc) radius region.

The HRC image resolves the narrow line region into X-ray emission
clumps matching bright emission-line clouds in the HST [OIII]
$\lambda$5007 images.  Two distinct X-ray knots are revealed at 1.4
arcsec NE and SW of the nucleus.  With image deconvolution technique,
we resolve sub-arcsec X-ray emission features tracing the collimation
and bending of the jet.

Based on the combined X-ray, [OIII], and radio continuum morphology,
we identify three locations of radio jet--cloud interaction.  The
[OIII] to soft X-ray ratios show that these clouds are strongly
affected by shock heating, whereas near two clouds the jet simply
thrust through with no signs of strong interaction.  This is further
strengthened by the presence of a $kT\sim 1$ keV collisionally ionized
component in the ACIS spectrum of the cloud G.  We further estimate
that the kinematic luminosity (or energy injection rate) of the
jet-driven shocks is $L_{K.E.}=6\times 10^{38}$ erg s$^{-1}$, which
shows a negligible amount of the jet power
($L_{K.E.}/P_{jet}=10^{-4}$) is lost to the ISM during the jet-cloud
interaction.

Our results show that the NGC 1068 radio jet plays a crucial role in
shaping the morphology, the kinematics and the ionization structure of
the innermost NLR.  Recent high spatial resolution studies based on
integral-field spectroscopy with Adaptive Optics in the near-IR start
to provide velocity fields and trace highly ionized gas in the radial
outflows in the NLR of Seyfert galaxies \citep[e.g.,][]{SB10,
Muller-Sanchez11}.  As shown in \citet{SB10}, [FeII] may be enhanced
where jet-induced shocks probably release the Fe locked in grains.  We
expect to see enhanced [FeII] peaks at the locations of jet-NLR
interaction suggested in this work.  In addition, the deep ACIS HETG
grating observations of NGC 1068 may provide a more detailed picture
of the kinematics and ionization of the hot phase ISM in NGC 1068.

\acknowledgments

We thank the anonymous referee for careful reading of the manuscript
and helpful comments. We thank Jack Gallimore for kindly providing the
VLA and MERLIN maps.  This work is supported by NASA grant GO1-12120X
(PI: Fabbiano).  We acknowledge support from the CXC, which is
operated by the Smithsonian Astrophysical Observatory (SAO) for and on
behalf of NASA under Contract NAS8-03060. This material is based upon
work supported in part by the National Science Foundation Grant
No. 1066293 and the hospitality of the Aspen Center for Physics.
J.W. acknowledges support from NASA grants GO1-12009X and
GO-12365.01-A.  M.K. is a member of CXC.  This research has made use
of data obtained from the {\em Chandra} Data Archive, and software
provided by the CXC in the application packages CIAO.

{\it Facilities:} \facility{CXO (HRC, ACIS)}

\clearpage

\begin{deluxetable}{ccccccccc}
\rotate
\tabletypesize{\scriptsize}
%\tabletypesize{\small}
\tablecaption{Measured X-ray and [OIII] Fluxes\label{flux}}
\tablewidth{0pt}
\tablehead{
\colhead{\begin{tabular}{c}
Cloud\\
label\\
\end{tabular}} &
\colhead{\begin{tabular}{c}
Distance\\
to Nuc. ($\arcsec$)\\
\end{tabular}} &
\colhead{\begin{tabular}{c}
Distance\\
to Nuc. (pc)\\
\end{tabular}} &
\colhead{\begin{tabular}{c}
R.A.=02$^{\rm h}$ 42$^{\rm m}$\\
(J2000)\\
\end{tabular}} &
\colhead{\begin{tabular}{c}
Decl.=$-$00$^{\rm o}$ 00$^{\prime}$\\
(J2000)\\
\end{tabular}} &
\colhead{\begin{tabular}{c}
Extraction\\
Area\\
\end{tabular}} &
\colhead{\begin{tabular}{c}
[OIII] flux\\
(c.g.s)\\
\end{tabular}} &
\colhead{\begin{tabular}{c}
0.5-2 keV flux\\
(c.g.s)\\
\end{tabular}} &
\colhead{\begin{tabular}{c}
[OIII]/soft X\\
\end{tabular}}
}
\startdata
B+C & 0.25 & 18 & 40.$^{\rm s}$719 & 47\farcs\/71 & 0\farcs\/4$\times$0\farcs\/45 & 15.4 & 16.6  &   $0.9\pm 0.1$    \\
D+E & 0.6 & 43 & 40.$^{\rm s}$724 & 47\farcs\/18 & 0\farcs\/4$\times$0\farcs\/4 &  14.2 & 1.9  &    $7.3\pm 0.6$    \\
F & 0.7 & 50 & 40.$^{\rm s}$752 & 47\farcs\/40 & 0\farcs\/35$\times$0\farcs\/35 & 14.0 & 1.5   &   $9.6\pm 1.0$     \\
G & 1.4 & 101 & 40.$^{\rm s}$758 & 46\farcs\/59 & 0\farcs\/45$\times$0\farcs\/4 & 7.7 & 2.9   &   $2.6\pm 0.2$      \\
H & 1.3 & 94 & 40.$^{\rm s}$675 & 49\farcs\/31 & 0\farcs\/45$\times$0\farcs\/4 & 0.73 & 0.38   &   $1.9\pm 0.9$     \\
\enddata
\tablecomments{The [OIII] cloud designations are adopted from Evans et al. (1991), except cloud H was not previously labelled and assigned here.  A uncertainty of $\sim$5\% for [OIII] flux is adopted (Capetti et al. 1997).  The [OIII] and X-ray fluxes are in unit of $10^{-12}$ erg s$^{-1}$ cm$^{-2}$ per arcsec$^2$.}
\end{deluxetable}

\clearpage

\begin{thebibliography}{}

\bibitem[Axon et al.(1998)]{Axon98} Axon, D.~J., Marconi, A., 
Capetti, A., et al.\ 1998, \apjl, 496, L75 

\bibitem[Bianchi et al.(2006)]{Bianchi06} Bianchi, S., Guainazzi, M., \& Chiaberge, M.\ 2006, \aap, 448, 499 

\bibitem[Bianchi et al.(2010)]{Bianchi10} Bianchi, S., Chiaberge, M., Evans, D.~A., et al.\ 2010, \mnras, 405, 553 

\bibitem[Bicknell et al.(1998)]{Bicknell98} Bicknell, G.~V., Dopita, M.~A., Tsvetanov, Z.~I., \& Sutherland, R.~S.\ 1998, \apj, 495, 680

\bibitem[Bland-Hawthorn et al.(1997)]{Bland-Hawthorn97} Bland-Hawthorn, J., Gallimore, J.~F., Tacconi, L.~J., et al.\ 1997, \apss, 248, 9 

\bibitem[Brinkman et al.(2002)]{Brinkman02} Brinkman, A.~C., Kaastra, J.~S., van der Meer, R.~L.~J., et al.\ 2002, \aap, 396, 761 

\bibitem[Capetti et al.(1997)]{Capetti97} Capetti, A., Axon, D.~J., \& Macchetto, F.~D.\ 1997, \apj, 487, 560 

\bibitem[Cavagnolo et al.(2010)]{Cavagnolo10} Cavagnolo, K.~W., McNamara, B.~R., Nulsen, P.~E.~J., et al.\ 2010, \apj, 720, 1066 

\bibitem[Crenshaw \& Kraemer(2000)]{Crenshaw00} Crenshaw, D.~M., \& Kraemer, S.~B.\ 2000, \apjl, 532, L101 

\bibitem[Esch et al.(2004)]{Esch04} Esch, D.~N., Connors, A., 
Karovska, M., \& van Dyk, D.~A.\ 2004, \apj, 610, 1213 

\bibitem[Exposito et 
al.(2011)]{Exposito11} Exposito, J., Gratadour, D., Cl{\'e}net, Y., \& Rouan, D.\ 2011, \aap, 533, A63 

\bibitem[Evans et al.(1991)]{Evans91} Evans, I.~N., Ford, 
H.~C., Kinney, A.~L., et al.\ 1991, \apjl, 369, L27 

\bibitem[Evans et al.(2010)]{Evans10} Evans, D.~A., Ogle, P.~M., Marshall, H.~L., et al.\ 2010, Accretion and Ejection in AGN: a Global View, 427, 97 

\bibitem[Ferland(2004)]{Ferland04} Ferland, G.~J.\ 2004, 
University of Kentucky Internal Report, 565 pages,  

\bibitem[Gallimore et al.(1996a)]{Gallimore96a} Gallimore, J.~F., 
Baum, S.~A., O'Dea, C.~P., \& Pedlar, A.\ 1996, \apj, 458, 136 

\bibitem[Gallimore et al.(1996b)]{Gallimore96b} Gallimore, J.~F., 
Baum, S.~A., \& O'Dea, C.~P.\ 1996, \apj, 464, 198 

\bibitem[Gallimore et al.(2004)]{Gallimore04} Gallimore, J.~F., 
Baum, S.~A., \& O'Dea, C.~P.\ 2004, \apj, 613, 794 

\bibitem[Groves et al.(2004)]{Groves04} Groves, B.~A., Cecil, 
G., Ferruit, P., \& Dopita, M.~A.\ 2004, \apj, 611, 786 

\bibitem[Karovska et al.(2005)]{Karovska05} Karovska, M., 
Schlegel, E., Hack, W., Raymond, J.~C., \& Wood, B.~E.\ 2005, \apjl, 623, L137 

\bibitem[Karovska et al.(2007)]{Karovska07} Karovska, M., Carilli, 
C.~L., Raymond, J.~C., \& Mattei, J.~A.\ 2007, \apj, 661, 1048 

\bibitem[Kinkhabwala et al.(2002)]{Kinkhabwala02} Kinkhabwala, A., et al.\ 2002, \apj, 575, 732

\bibitem[Kishimoto(1999)]{Kishimoto99} Kishimoto, M.\ 1999, \apj, 518, 676 

\bibitem[Krips et al.(2011)]{Krips11} Krips, M., Mart{\'{\i}}n, 
S., Eckart, A., et al.\ 2011, \apj, 736, 37 

\bibitem[Li et al.(2003)]{Li03} Li, J., Kastner, J.~H., 
Prigozhin, G.~Y., \& Schulz, N.~S.\ 2003, \apj, 590, 586 

\bibitem[Matt et al.(2000)]{Matt00} Matt, G., Fabian, A.~C., 
Guainazzi, M., et al.\ 2000, \mnras, 318, 173 

\bibitem[M{\"u}ller S{\'a}nchez et al.(2009)]{Muller-Sanchez09} M{\"u}ller S{\'a}nchez, F., Davies, R.~I., Genzel, R., et al.\ 2009, \apj, 691, 749

\bibitem[M{\"u}ller-S{\'a}nchez et al.(2011)]{Muller-Sanchez11} M{\"u}ller-S{\'a}nchez, F., Prieto, M.~A., Hicks, E.~K.~S., et al.\ 2011, \apj, 739, 69 

\bibitem[Murray et al.(2000)]{Murray00} Murray, S.~S., Austin, G.~K., Chappell, J.~H., et al.\ 2000, \procspie, 4012, 68 

\bibitem[Ogle et al.(2003)]{Ogle03} Ogle, P.~M., Brookings, T., Canizares, C.~R., Lee, J.~C., \& Marshall, H.~L.\ 2003, \aap, 402, 849 

\bibitem[Paggi et al.(2012)]{Paggi12} Paggi, A., Wang, J., Fabbiano, G., Elvis, M., \& Karovska, M.\ 2012, ApJ in press, arXiv:1203.1279 

\bibitem[Raban et al.(2009)]{Raban09} Raban, D., Jaffe, W., R{\"o}ttgering, H., Meisenheimer, K., \& Tristram, K.~R.~W.\ 2009, \mnras, 394, 1325 

\bibitem[Rodr{\'{\i}}guez-Ardila et al.(2006)]{Rodriguez06} 
Rodr{\'{\i}}guez-Ardila, A., Prieto, M.~A., Viegas, S., 
\& Gruenwald, R.\ 2006, \apj, 653, 1098 

\bibitem[Storchi-Bergmann et al.(2010)]{SB10}  Storchi-Bergmann, T., Lopes, R.~D.~S., McGregor, P.~J., Riffel, R.~A.,  Beck, T., \& Martini, P.\ 2010, \mnras, 402, 819 

\bibitem[Storchi-Bergmann et al.(2012)]{SB12} 
Storchi-Bergmann, T., Riffel, R.~A., Riffel, R., et al.\ 2012, ApJ in press, arXiv:1206.4014 

\bibitem[Wang et al.(2009a)]{Wang09a} Wang, J., Fabbiano, G., 
Karovska, M., et al.\ 2009, \apj, 704, 1195 


\bibitem[Wang et al.(2009b)]{Wang09b} Wang, J., Fabbiano, G., 
Elvis, M., et al.\ 2009, \apj, 694, 718 

\bibitem[Wang et al.(2011a)]{Wang11a} Wang, J., Fabbiano, G., 
Elvis, M., et al.\ 2011, \apj, 736, 62 


\bibitem[Wang et al.(2011b)]{Wang11b} Wang, J., Fabbiano, G., 
Risaliti, G., et al.\ 2011, \apj, 729, 75 

\bibitem[Wilson 
\& Ulvestad(1983)]{Wilson83} Wilson, A.~S., \& Ulvestad, J.~S.\ 1983, \apj, 275, 8 

\bibitem[Young et al.(2001)]{Young01} Young, A.~J., Wilson,
A.~S., \& Shopbell, P.~L.\ 2001, \apj, 556, 6

\bibitem[Zacharias et al.(2004)]{Zacharias04} Zacharias, N., Monet, 
D.~G., Levine, S.~E., et al.\ 2004, Bulletin of the American Astronomical 
Society, 36, 1418 

\end{thebibliography}
\end{document}